%Paper: hep-th/9505088
%From: Matt Strassler <strasslr@physics.rutgers.edu>
%Date: Tue, 16 May 1995 01:30:03 -0400

%
\input harvmac
%
% defs to allow sub-sub-sections; Leigh (13/5/95)
%
\global\newcount\subsubsecno \global\subsubsecno=0
\def\subsec#1{\global\advance\subsecno by1\global\subsubsecno=0
\message{(\secsym\the\subsecno. #1)}
\ifnum\lastpenalty>9000\else\bigbreak\fi
\noindent{\it\secsym\the\subsecno. #1}\writetoca{\string\quad
{\secsym\the\subsecno.} {#1}}\par\nobreak\medskip\nobreak}
\def\subsubsec#1{\global\advance\subsubsecno by1
\message{(\secsym\the\subsecno.\the\subsubsecno. #1)}
\ifnum\lastpenalty>9000\else\bigbreak\fi
{\it\secsym\the\subsecno.\the\subsubsecno. #1}\writetoca{\string\quad
{\secsym\the\subsecno.\the\subsubsecno.} {#1}}\par\nobreak\medskip\nobreak}
%
% def for multi-line titles; Leigh (14/3/95)
%
\def\MTitle#1#2#3{\nopagenumbers\abstractfont\hsize=\hstitle\rightline{#1}%
\vskip 1in\centerline{\titlefont #2}\vskip .08in\centerline{\titlefont #3}
\abstractfont\vskip .2in\pageno=0}
%
% refs defs
%
\def\PLB#1{Phys. Lett. {\bf #1B}}
\def\PRD#1{Phys. Rev. {\bf D#1}}
\def\NPB#1{Nucl. Phys. {\bf B#1}}

\def\npb#1#2#3{Nucl. Phys. {\bf B#1} (#2) #3}

\nref\Holo{N. Seiberg, \PLB{318} (1993) 469; {\it The Power of
Holomorphy: Exact Results in 4-D SUSY Field Theories.} (hep-th/9408013).}

\nref\Seibone{N. Seiberg, \PRD{49} (1994) 6857,  (hep-th/9402044).}

\nref\ILS{K. Intriligator, R.G. Leigh and N. Seiberg, \PRD{50} (1994) 1092,
(hep-th/9403198).}

\nref\NSEW{N. Seiberg and E. Witten, \NPB{426} (1994) 19;
\NPB{430} (1994) 485; \NPB{431} (1994) 484.}

\nref\NAD{N. Seiberg, \NPB{435} (1995) 129;
{\it Electric-Magnetic Duality in Supersymmetric
Non-Abelian Gauge Theories}, hep-th/9411149.}

\nref\KUT{D. Kutasov, {\it A Comment on Duality in N=1 Supersymmetric
Non-Abelian Gauge Theories}, EFI--95--11, hep-th/9503086.}

\nref\emop{R.G. Leigh and M.J. Strassler,
{\it Exactly Marginal Operators and Duality in
Four Dimensional N=1 Supersymmetric Gauge Theory},
RU--95--2, hep-th/9503121.}

\nref\somods{K. Intriligator and N. Seiberg, {\it Duality, Monopoles, Dyons,
Confinement and Oblique Confinement in Supersymmetric $SO(\nc)$ Gauge
Theories}, RU--95--3, hep-th/9503179.}

\nref\asy{O. Aharony, J. Sonnenschein and S. Yankielowicz, {\it Flows
and Duality Symmetries in N=1 Supersymmetric Gauge Theories}, TAUP--2246--95,
CERN-TH/95--91, hep-th/9504113.}

\nref\kutschw{D. Kutasov and A. Schwimmer, {\it On Duality in Supersymmetric
Yang-Mills Theory}, EFI--95--20, WIS/4/95, hep-th/9505004.}

\nref\kenpoul{K. Intriligator and P. Pouliot, {\it Exact Superpotentials,
Quantum Vacua and Duality in Supersymmetric $Sp(\nc)$ Gauge Theories},
RU-95-23, hep-th/9505006.}

\nref\keni{K. Intriligator, {\it New RG Fixed Points and Duality in
Supersymmetric $Sp(\nc)$ and $SO(\nc)$ Gauge Theories}, RU--95--27,
hep-th/9505051.}

\nref\nistwo{A. Klemm, W. Lerche,  S.
Theisen and S. Yankielowicz, \PLB{344} (1995) 169 (hep-th/9411048);
P.C. Argyres and A.E. Faraggi, {\it The Vacuum
Structure and Spectrum of N=2 Supersymmetric $SU(N)$
Gauge Theory}, IASSNS-HEP-94-94, hep-th/9411057;
M.R. Douglas and  S.H. Shenker, {\it Dynamics
of $SU(N)$ Supersymmetric Gauge Theory}, RU-95-12, hep-th/9503163;
P.C. Argyres and M.R. Douglas, {\it New Phenomena
in $SU(3)$ Supersymmetric Gauge Theory}, RU-95-31, hep-th/9505062.}

\nref\others{O. Aharony, {\it Remarks on Non-abelian Duality in
N=1 Supersymmetric Gauge Theories}, TAUP-2232-95, hep-th/9502013;
S. Elitzur, A. Forge, A. Giveon and E. Rabinovici,
{\it More results in N=1 Supersymmetric Gauge Theories},
RI-4-95 hep-th/9504080.}

\nref\ads{I. Affleck, M. Dine, and N. Seiberg, \npb{241}{1984}{493};
\npb{256}{1985}{557}}%

\nref\spa{P. Argyres, private communication.}

% macros
%
% pick a letter for the adjoint
\def\ad{X}
\def\adD{Y}

\font\zfont = cmss10 %scaled \magstep1
\def\ZZ{\hbox{\zfont Z\kern-.4emZ}}
\def\bigone{\hbox{1\kern-.2eml}}

\def\vl{p}

% Accents

%
% Irreps
\def\nc{{N_{c}}}\def\nf{{N_{f}}}\def\ncd{{\tilde\nc}}
% gauge trace
\def\tr#1{{\rm tr} #1}
% flavor trace
\def\Tr#1{{\rm Tr} #1}
%
% Fields

\def\lk#1{\lambda_{#1}}

\def\tQ{\tilde{Q}}

\MTitle{hep-th/9505088, RU-95-30}
{\vbox{\centerline{Duality of $Sp(2\nc)$ and $SO(\nc)$}}}
{\vbox{\centerline{Supersymmetric Gauge Theories with Adjoint Matter}}}
\bigskip
\centerline{R.G. Leigh$^{\dagger,\ddagger}$ and M.J. Strassler$^\dagger$}
\bigskip
\centerline{\it $^\dagger$Department of Physics and Astronomy}
\baselineskip=14pt
\centerline{\it Rutgers University}
\baselineskip=14pt
\centerline{\it Piscataway, NJ 08855-0849}
\bigskip\smallskip
\centerline{\it $^\ddagger$CERN, TH Division}
\baselineskip=14pt
\centerline{\it CH-1211 Geneva 23}
\baselineskip=14pt
\centerline{\it Switzerland}
\vglue 0.8cm
\centerline{\rm ABSTRACT}
\vglue 0.3cm
\bigskip
\baselineskip 18pt
\noindent

We discuss electric-magnetic duality in two new classes of supersymmetric
Yang-Mills theories. The models have gauge group $Sp(2\nc)$ or $SO(\nc)$
with matter in both the adjoint and defining representations.
By perturbing these theories with various superpotentials, we find a
variety of new infrared fixed points with dual descriptions. This
work is complementary to that of Kutasov and Schwimmer on $SU(\nc)$
and of Intriligator on other models involving $Sp(2\nc)$ and $SO(\nc)$.

\Date{5/95}
%\draftmode

%Supported in part by the U.S. Department of Energy, contract
%%DE-FG05-90ER40599.

\newsec{Introduction}

Recent developments in the study of supersymmetric Yang-Mills theories
have led to the discovery of a wide variety of phenomena and have given
us insight into many aspects of strongly coupled physics
\refs{\Holo-\nistwo}.
Among these phenomena are large classes of superconformally
invariant infrared fixed points \NAD. In recent months, the study of
these four-dimensional theories has led to a realization of their
rich structure.

Perhaps the most exciting property of supersymmetric Yang-Mills
theories is that they can be shown to have a form of electric-magnetic
duality \refs{\NSEW-\keni}.
The study of duality in N=1 models has just begun ---
only a few simple classes of models have been investigated. It is
important to explore other possibilities; with
the elucidation of many dual models, a pattern may become apparent,
leading eventually to an understanding of the general principles involved.

In this paper, we consider Yang-Mills theories with gauge
group $Sp(2\nc)$ and $SO(\nc)$ with matter in the adjoint
representation as well as the defining representation.
These models are very similar to those recently studied in
Refs.~\refs{\KUT, \kutschw} and are  complementary to those
studied in Ref.~\keni. Because of these similarities, we follow closely
the exposition given in \kutschw.  In our conclusion we comment
on possible relations between the duality in these models
and that of finite N=2 theories \refs{\NAD, \emop, \kutschw}.

\newsec{$Sp(2\nc)$ with an adjoint and fundamentals}

\subsec{Preliminaries}

We consider an $Sp(2\nc)$ supersymmetric Yang-Mills theory\foot{Note that
Refs.~{\keni,\NAD,\kenpoul} refer to this group as $Sp(\nc)$.}
with an adjoint field $\ad$ (which may be taken as a symmetric
2-tensor) and $2\nf$ fundamentals $Q^i$. The coefficient of the
one-loop $\beta$-function
is $b_0=2(\nc+1)-\nf$. For some values of $\nf$, we expect that
at the origin of moduli space there is an interacting conformal
field theory in the infrared, which may possess dual descriptions.
As yet, we do not understand the duality of this model for arbitrary
superpotential. However, the description is straightforward if
we include a superpotential
\eqn\wspe{W= \lk{k+1} {1\over 2(k+1)}\tr \ad^{2(k+1)}.}
Traces are taken with the invariant tensor $J$ which may be thought
of as the matrix $\bigone_{\nc}\otimes i\sigma_2$.
The flavor group of these models is $SU(2\nf)\times U(1)_R$; the matter
fields transform as:
\eqn\transQ{\eqalign{
Q\in\left( {\bf 2\nf}, 1-{\nc+1\over\nf(k+1)}\right)\cr
\ad\in\left( {\bf1}, {1\over k+1}\right).}}
We may form the following gauge invariant polynomials in these fields:
\eqn\gifsp{\eqalign{
u_n =&\; {1\over 2n}\tr \ad^{2n}\cr
M^{rs}_{(2n)}=&\; Q^r\ad^{2n}Q^s\cr
M^{rs}_{(2n+1)}=&\; Q^r\ad^{2n+1}Q^s}}
where $n=0,1,2,\ldots$. The mesons $M_{2n}$ transform under the
$\nf(2\nf-1)$-dimensional antisymmetric tensor representation of
$SU(2\nf)$ while $M_{2n+1}$ resides in the $\nf(2\nf+1)$-dimensional
symmetric tensor representation. Of course, many of these are
redundant operators. In the presence of the superpotential \wspe,
the chiral ring consists of the operators $u_n$ and $M_{(2n)}$
for $n=0,1,\ldots,k$ and $M_{(2n+1)}$ for $n=0,1,\ldots,k-1$.

\subsec{Stability}
We wish to study under what circumstances this model possesses stable
ground states. Consider deforming the model \wspe\ to
\eqn\defsp{W=\sum_{n=1}^{k+1} \lk{n}u_n.}
The field $X$ can be rotated into the $2\times 2$ block form
$\ad={\rm diag}(x_1\sigma_1,x_2\sigma_1,\ldots)$, where $\sigma_1$
is a Pauli matrix.
Given such a superpotential, one finds supersymmetric
ground states for $\langle Q\rangle=0$ and $x_i$ satisfying the
equation
\eqn\WprimeSp{x\sum_{n=0}^k \lk{n+1}\; x^{2n}
              = x\prod_{j=1}^k (x^2-\alpha_{(j)}^2) = 0.
}
Generically there are $k+1$ independent solutions of this equation
(the overall sign of an eigenvalue is unobservable)
which we will label $\alpha_{(j)}$, $j=0,\ldots,k$, with $\alpha_{(0)}=0$.
A ground state will then
be labeled by a set of integers $(\vl_0,\vl_1,\ldots,\vl_{k})$;
each $\vl_j$ gives the number of eigenvalues $x_i$ of $\vev{X}$ which
are equal to $\alpha_{(j)}$.
The symmetry surviving in such a vacuum state is
\eqn\symvacsp{ Sp(2\vl_0)\times U(\vl_1)\times\ldots\times U(\vl_{k})}
with $\sum_{j=0}^{k}\vl_j=\nc$. In these vacua, the adjoint fields will
be massive and can be integrated out.  The
matter fields $Q$ decompose as follows: there are $2\nf$ flavors of fields
in the fundamental and antifundamental representations of
each of the unitary group factors, while there remain $\nf$ flavors
($2\nf$ fields $Q^r$)
in the symplectic factor. The resulting low-energy theory
is thus a product of SQCD-like models; these are known to possess
stable vacua  for $2\nf \geq \vl_j$, $j>0$ \ads\
and $\nf \geq \vl_0+1$ \kenpoul.
We conclude that the theory with
superpotential \defsp\ has stable vacua for $\nf\geq (\nc+1)/(2k+1)$;
the same is
true for the theory with superpotential \wspe, assuming the result
varies smoothly for small coupling $\lk{n}$, $n=1,\ldots, k$.

If the coefficients $\lk{n}$ are properly tuned, some of the solutions
$\alpha_{(j)}$ will coincide.  In this case the $x_i$ satisfy the
equation
\eqn\WprimeSpB{ x^{1+2r_0}\prod_{j=1}^m (x^2-\alpha_{(j)}^2)^{r_j} = 0.
}
such that $\sum r_j = k$.  The $Sp(2\vl_0)$ group factor now
has a massless adjoint superfield with
an effective superpotential $\Tr X^{2(r_0+1)}$; similarly the $U(\vl_j)$
factor corresponding to the $\vl_j$ eigenvalues $x_i = \alpha_{(j)}$
has an adjoint superfield with
effective superpotential $\Tr X^{r_j+1}$.  Such $U(\nc)$
theories were studied in \kutschw.

\subsec{The dual model}

We will refer to the above $Sp(2\nc)$ model as the electric theory.
The dual magnetic theory is an $Sp(2\ncd)$ gauge theory with
$\ncd=(2k+1)\nf-\nc-2$. Under the duality transformation, the
mesons $M_{(2n)}$ and $M_{(2n+1)}$ of \gifsp\ are mapped to gauge
singlet fields which we will refer to by the same name. In addition
the magnetic theory possesses $2\nf$ fields $q$ in the fundamental
representation of $Sp(2\nc)$ and an adjoint field $Y$. The superpotential
of the magnetic theory is:
\eqn\wspm{W_m= \lk{k+1} {1\over 2(k+1)}\;\tr \adD^{2(k+1)}
+\sum_{n=0}^{2k} M^{rs}_{(n)}\; q_r\adD^{2k-n} q_s.}

The 't Hooft anomaly matching conditions for the global symmetries
at the origin of moduli space are powerful constraints.
The fields of the magnetic theory transform
as follows under the $SU(2\nf)\times~U(1)_R$ flavor symmetry:
$$q\in\left({\bf \overline{2\nf}},\;1-{\ncd+1\over\nf(k+1)}\right),$$
$$\adD\in\left({\bf 1},\; {1\over k+1}\right),$$
$$M_{(2n)}\in\left({\bf\nf(2\nf-1)},\;
2-{2(\nc+1)-2n\nf\over\nf(k+1)}\right),$$
$$M_{(2n+1)}\in\left({\bf\nf(2\nf+1)},\;
2-{2(\nc+1)-(2n+1)\nf\over\nf(k+1)}\right).$$
The anomalies are
found to be, in both the electric and magnetic theories:
\eqn\anomsp{\eqalign{
U(1)_R:&\;\; -{\nc(2\nc+3)\over k+1}\cr
U(1)_R^3:&\;\; -4{\nc(\nc+1)^3\over\nf^2 (k+1)^3}+\nc(2\nc+1)
\left( 1-\left( {k\over k+1}\right)^3\right)\cr
U(1)_R SU(2\nf)^2:&\;\; -{\nc(\nc+1)\over\nf (k+1)}\cr
SU(2\nf)^3:&\;\; +2\nc . }}

\subsec{Deformations}

We first consider the addition of a mass term for a quark in the
electric theory. This perturbation will cause flow to a similar
theory with one fewer flavor of quarks. In the dual theory, we expect,
following the duality discussed above, to flow to a theory with
gauge group $Sp(2[\ncd-2k-1])$. Indeed, the mass perturbation
$mQ^{2\nf-1}Q^{2\nf}$ gives rise to a superpotential for the
magnetic theory of the form:
$$W_m=\lk{k+1} {1\over 2(k+1)}\;\tr \adD^{2(k+1)}
+\sum_{n=0}^{2k} M^{rs}_{(n)}\; q_r\adD^{2k-n} q_s + mM_{(0)}^{2\nf-1,2\nf}$$
The $M$ equations of motion then imply the vacuum expectation values satisfy:
\eqn\vacmass{\langle q_{2\nf-1} \adD^{j}q_{2\nf}\rangle=-m\delta_{j,2k}\;
;\; j=0,1,\ldots,2k}
The $D$-flatness and other $F$-flatness conditions result in
expectation values which
break $Sp(2\ncd)\rightarrow Sp(2[\ncd-2k-1])$,
as required.  The fields $q_{2\nf-1},q_{2\nf}$ are eaten by the broken
gauge multiplets.  The adjoint field $Y$ decomposes into an adjoint
of the low energy gauge group, along with $2k+1$ flavors $Z_a,Z'_a$,
of which $2k$ are eaten by gauge multiplets; the remaining one becomes
massive through the $\Tr Y^{2(k+1)}\rightarrow Z_1\vev{Y^{2k}}Z'_1$
interaction. In the end, the number of flavors remaining is $\nf-1$.
(Some massive singlets are also generated.)

We can also study the effects of adding
$Q^{2\nf-1}\ad^{j}Q^{2\nf}$ for $j=1,\ldots,2k$
to the superpotential of the electric theory. In this case,
the gauge group of the magnetic theory is reduced by $2k-j+1$ colors;
the fields  $q_{2\nf-1}, q_{2\nf}$ are eaten by the broken gauge
multiplets;
the field $Y$ decomposes into an adjoint of the low energy gauge group and
$2k-j+1$ new flavors $Z_a,Z'_a$,
of which $2k-j$ are eaten by gauge multiplets and the remaining one
develops an interaction $Z_1Y^{j}\vev{Y^{2k-j}}Z'_1$ from the
$\Tr Y^{2(k+1)}$ interaction. In the end, the number of flavors
remaining is $\nf-1$, with an extra flavor coupled to the adjoint through a
Tr $Z_1Y^{j}Z'_1$ term.
This corresponds to the type of duality studied in
Ref.~\asy\ in the case of an $SU(\nc)$ gauge group.\foot{
Thus the duality of Ref.~\asy, along with many generalizations,
may be straightforwardly derived from that of Refs.~{\KUT,\kutschw.}}

Lastly, we may consider perturbations by operators of the form
$u_j$. We have given the general analysis of this situation in Section
2.2. A particularly simple example is $j=1$, corresponding to a mass
term for the field $\ad$.   Let us first discuss the electric
theory.  For simplicity, we consider the case $\nf\geq\nc-1.$
In the presence of the
mass term for $\ad$, the eigenvalues of $\ad$ must satisfy
$$x(m+\lk{k}x^{2k})=0$$
The solutions, $x=0,\eta_i(-m/\lk{k+1})^{1/2k}$, with $\eta_i^{2k}=1$,
constitute $k+1$ independent values.  At scales well below
the adjoint mass, we have an
$$Sp(2\vl_{0})\times U(\vl_{1})\ldots\times U(\vl_{k})$$
gauge symmetry, as was explained in Section 2.2, with
$\sum_{r=0}^{k}\vl_r=\nc$.  All the vacua are stable.
In the dual theory, we also have a massive adjoint field, and the
analysis proceeds similarly. However, in this case, in identifying
the ground states, we must take care to check stability. It is
convenient to write the symmetry group of the magnetic theory in
the form:
$$Sp(2[(2k+1)\nf-\nc-2])\rightarrow Sp(2[\nf-j_0-2])\times
U(2\nf-j_1)\times\ldots\times U(2\nf-j_k)$$
where $\sum_{r=0}^k j_r=\nc$.  Each of these groups contains
the appropriate magnetic $Mqq$ coupling, where the singlet $M$ is a linear
combination of the $M_{(j)}$ of the unbroken magnetic theory.
If $j_0=-1$, then the $Sp[2(\nf+1)]$ theory confines and has no
vacuum \keni.\foot{The field $N_{rs}=(q_rq_s)$ has superpotential
$W=M^{rs}N_{rs}$ and a constraint
${\rm Pf} N = \Lambda_L^{2(\nc+1)}$; these are inconsistent.}
If $j_r=0$, $r>0$, then the $SU(2\nf)$
theory confines but possesses a vacuum in
which its baryons $B,\tilde B$ condense and
break the remaining $U(1)$\kutschw.
\foot{The field $N_r^s=(q_r\tilde q^s)$ has superpotential
$W=M^r_sN_r^s$ and a constraint
${\rm det} N  - B\tilde B= \Lambda_L^{2\nc}$; these are consistent.}
Thus, stability requires $j_r\geq 0$ for all $r$
and so there are the same number of  stable ground states as
in the electric theory.
With $p_r$ and $j_r$ identified, it becomes
clear that this duality is consistent with that discovered in \NAD\ and
explored further in \kenpoul.

It is interesting to consider a few specific examples.
Suppose $\vl_0 = 0$.  Then the electric gauge group is
a product of $U(\vl_i)$ factors.  The magnetic group is
$Sp(2[\nf-2])$ times a product of $U(2N_f-\vl_i)$ factors.
The $Sp(2[\nf-2])$ confines \kenpoul\ and its mesons $N=qq$ acquire
mass through the coupling $Mqq$ in the magnetic superpotential.
The low energy magnetic $Sp(2[\nf-2])$ theory is therefore empty.

Conversely, if $\vl_0 =\nf-2$, then the electric $Sp(2[\nf-2])$ factor
confines into mesons $M=QQ$ and develops a superpotential
${\rm Pf} M$ \kenpoul.  The dual magnetic group is a product of
$U(2N_f-\vl_i)$ factors, but a number of singlet meson fields are left
over.  An instanton effect in the broken magnetic
$Sp$ gauge group generates
a term proportional to ${\rm Pf} M$ \NAD, \kenpoul\ and preserves
the duality.

Another example is when $\vl_0=\nc$; the field $\ad$ becomes
massive but the $Sp(2\nc)$ gauge group is unbroken.  In this case
the magnetic theory has a factor $Sp(2[\nf-\nc-2])$ and several factors
of $U(2\nf)$, each of which has $2\nf$ flavors.  As discussed in
\kutschw, the  $SU(2\nf)$ theories confine and their baryons
condense, breaking the $U(1)$ groups.  The magnetic theory reduces
to the $Sp(2[\nf-\nc-2])$ theory expected from \refs{\NAD, \kenpoul}.

One may also consider more general deformations as in Eq.~\defsp.
For generic values of the couplings, the analysis is similar
to the above.
At special values of the couplings, roots of $W'(x)=0$ may coincide,
leading to low energy theories which have adjoint matter and
superpotentials such as $\Tr X^{r}$.  As an example, consider
the superpotential with $W'(x) = x^{(2r+1)}(x^2-a^2)^{k-r}$.  The low
energy electric theory is
$$
Sp(2\vl_0)\times U(\nc-\vl_0)\
$$
with superpotential $\Tr X_{Sp}^{2(r+1)} + \Tr X_U^{k-r+1}$.
The magnetic theory has a low energy theory
$$
Sp(2\tilde\vl_0)\times U(\tilde\nc-\tilde\vl_0)\
$$
and the same superpotential as the electric theory.  But stability requires
that
$$
\nf \geq{(\tilde\nc-\tilde\vl_0)\over 2(k-r+1)}  \ ;
\nf \geq{(\tilde\vl_0+1)\over (2r+1)}  \ ;
$$
the former relation follows from \kutschw\ and the latter
from Section 2.2.  These force
$$(2r+1)\nf-\nc-2\leq\tilde\vl_0\leq(2r+1)\nf-1\ ,$$
showing that both theories have $\nc+1$ vacua.  It can easily be checked
that the duality illustrated
here and in \kutschw\ is maintained by these vacua.

Thus, in a fashion very similar to
that illustrated in Ref.~\kutschw, the deformations $u_j$ lead in general
to similar theories with lower values of $k$. The duality map is
preserved under these deformations.

\newsec{$SO(\nc)$ with an adjoint and vectors}

\subsec{Preliminaries}

We now consider $SO(\nc)$ super Yang Mills theory
with an adjoint field $\ad$ (which may be thought of as the anti-symmetric
2-tensor) and $\nf$ vectors $Q^i$. The one-loop $\beta$-function
is $b_0=2(\nc-2)-\nf$. We will again include a superpotential
\eqn\wspe{W_e= \lk{k+1} {1\over 2(k+1)}\tr \ad^{2(k+1)}.}
Traces are taken with the Kronecker $\delta$.
The flavor group of these models is $SU(\nf)\times U(1)_R$; the matter
fields transform as:
\eqn\transQso{\eqalign{
Q\in\left( {\bf \nf}, 1-{\nc-2\over\nf(k+1)}\right)\cr
\ad\in\left( {\bf1}, {1\over k+1}\right).}}
We may form the following gauge invariant polynomials in these fields:
\eqn\gifso{\eqalign{
u_n =&\; {1\over 2n}\tr \ad^{2n}\cr
M^{rs}_{(2n)}=&\; Q^r\ad^{2n}Q^s\cr
M^{rs}_{(2n+1)}=&\; Q^r\ad^{2n+1}Q^s}}
where $n=0,1,2,\ldots$. The mesons $M_{2n}$ transform under the
$\nf(\nf+1)/2$-dimensional symmetric tensor representation of
$SU(\nf)$ while $M_{2n+1}$ resides in the $\nf(\nf-1)/2$-dimensional
anti-symmetric tensor representation. In addition there are the generalized
baryon operators obtained by contracting fields with an $\epsilon$
tensor; we will not discuss them here.

Of course, many of these operators are
redundant. In the presence of the superpotential \wspe,
the chiral ring consists of the operators $u_n$ and $M_{(2n)}$
for $n=0,1,\ldots,k$ and $M_{(2n+1)}$ for $n=0,1,\ldots,k-1$, along
with a number of generalized baryon operators.

\subsec{Stability}

We wish to study under what circumstances this model possesses stable
ground states. Much of this analysis is carried over with little
modification from the previous sections.
Consider deforming the model \wspe\ to
\eqn\defso{W=\sum_{j=1}^{k+1} \lk{j}u_j.}
The field $X$ can be rotated into the $2\times 2$ block form
$\ad={\rm diag}(x_1i\sigma_2,
\ldots,x_{n_c}i\sigma_2)$ for $SO(2n_c)$ and $\ad={\rm diag}(x_1i\sigma_2,
\ldots,x_{n_c}i\sigma_2,0)$ for $SO(2n_c+1)$, where $\sigma_2$
is a Pauli matrix.
Given such a superpotential, one finds supersymmetric
ground states for $\langle Q\rangle=0$ and $x_i$ satisfying the
equation
$$x\sum_{n=0}^k \lk{n}\; x^{2n} = x\prod_{j=1}^k (x^2-\alpha_{(j)}^2) = 0.$$
Generically there are $k+1$ independent solutions of this equation
which we will label $\alpha_{(j)}$, $j=0,\ldots,k$, with $\alpha_{(0)}=0$.
A ground state will then
be labeled by a set of integers $(\vl_0,\vl_1,\ldots,\vl_{k})$;
each $\vl_j$ gives the number of eigenvalues $x_i$ of $\vev{X}$ which
are equal to $\alpha_{(j)}$.
The symmetry surviving in such a vacuum state is
\eqn\symvacsp{ \eqalign{SO(2n_c+1)\rightarrow SO(2\vl_0+1)&\times
U(\vl_1)\times\ldots\times U(\vl_{k})\cr
&{\rm or}\cr
SO(2n_c)\rightarrow SO(2\vl_0)\times & U(\vl_1)\times\ldots\times U(\vl_{k})
}}
with $\sum_{j=0}^{k}\vl_j=n_c$. In these vacua, the adjoint fields will
be massive and can be integrated out; the
matter fields $Q$ decompose into $\nf$ flavors in the fundamental
and anti-fundamental representations of
each of the unitary group factors, plus $\nf$ vectors
in the orthogonal group factor. The resulting low-energy theory
is thus a product of SQCD-like models; the unitary factors have
stable vacua for $\nf \geq \vl_j$, $j>0$ \ads\ while $SO(\nc)$ has
stable vacua for $\nf \geq \nc-4$ \NAD, \somods.
We conclude that the theory with
superpotential \defso\ has stable vacua for $\nf\geq (\nc-4)/(2k+1)$;
the same is true for the theory with superpotential \wspe, assuming the
result varies smoothly for small couplings $\lk{n}$, $n=1,\ldots, k$.

Again, when eigenvalues $\alpha_{(j)}$ coincide the low energy
theory has massless adjoint matter with a superpotential.  We omit
the details, since they are completely analogous to the previous
case.

\subsec{The dual model}

The above $SO(\nc)$ model will be referred to as the electric theory.
The dual magnetic theory is an $SO(\ncd)$ gauge theory with
$\ncd=(2k+1)\nf-\nc+4$. Under the duality transformation, the
mesons $M_{(2n)}$ and $M_{(2n+1)}$ of \gifsp\ are mapped to gauge
singlet fields which we will refer to by the same name. In addition
the magnetic theory possesses $\nf$ fields $q$ in the vector
representation of $SO(\nc)$ and an adjoint field $Y$. The superpotential
of the magnetic theory is:
\eqn\wspm{W_m= \lk{k+1} {1\over 2(k+1)}\;\tr \adD^{2(k+1)}
+\sum_{n=0}^{2k} M^{rs}_{(n)}\; q_r\adD^{2k-n} q_s.}

Let us check the 't Hooft anomaly matching conditions for the global symmetries
at the origin of moduli space.
The fields of the magnetic theory transform
as follows under the $SU(\nf)\times~U(1)_R$ flavor symmetry:
$$q\in\left({\bf \overline{N}_f},\;1-{\ncd-2\over\nf(k+1)}\right),$$
$$\adD\in\left({\bf 1},\; {1\over k+1}\right),$$
$$M_{(2n)}\in\left({\bf\nf(\nf+1)/2},\;
2-{2(\nc-2)-2n\nf\over\nf(k+1)}\right),$$
$$M_{(2n+1)}\in\left({\bf\nf(\nf-1)/2},\;
2-{2(\nc-2)-(2n+1)\nf\over\nf(k+1)}\right).$$
The anomalies are
found to be, in both the electric and magnetic theories:
\eqn\anomsp{\eqalign{U(1)_R:&\;\; -{\nc(\nc-3)\over 2(k+1)}\cr
U(1)_R^3:&\;\; -{\nc(\nc-2)^3\over\nf^2 (k+1)^3}+{\nc(\nc-1)\over2}
\left( 1-\left( {k\over k+1}\right)^3\right)\cr
U(1)_RSU(\nf)^2:&\;\; -{\nc(\nc-2)\over\nf (k+1)}\cr
SU(\nf)^3:&\;\; +\nc . }}

\subsec{Deformations}

We first consider the addition of a mass term for a quark in the
electric theory. This perturbation will cause flow to a similar
theory with one fewer flavor of quarks. The mass perturbation
$mQ^{\nf}Q^{\nf}$ gives rise to a superpotential for the
magnetic theory of the form
$$W_m=\lk{k+1} {1\over 2(k+1)}\;\tr \adD^{2(k+1)}
+\sum_{n=0}^{2k} M^{rs}_{(n)}\; q_r\adD^{2k-n} q_s + mM_{(0)}^{\nf,\nf}$$
The $M$ equations of motion then imply the vacuum expectation values satisfy:
\eqn\vacmass{\langle q_{\nf} \adD^{j}q_{\nf}\rangle=-m\delta_{j,2k}\;
;\; j=0,1,\ldots,2k}
The resulting expectation values for the individual fields
break $SO(\ncd)\rightarrow SO(\ncd-2k-1)$
as required, and there are $\nf-1$ massless vectors remaining.

The addition of operators $Q\ad^{j}Q$ for $j=1,\ldots,2k$
to the superpotential of the electric theory, in precise analogy
to the symplectic case discussed above, causes
the gauge group of the magnetic theory to be
reduced and a certain superpotential
to be induced in the low energy magnetic theory.

Now consider perturbations by operators of the form
$u_j$. We have given the general analysis of this situation in Section
3.2. For $j=1$ the eigenvalues of $\ad$ must satisfy
$$x(m+\lk{k}x^{2k})=0$$
as in the $Sp(2\nc)$ case. At scales well below
the adjoint mass, we have an
$$SO\left(\nc-2\sum_{r=1}^k \vl_r\right)
\times U(\vl_{1})\ldots\times U(\vl_{k})$$
gauge symmetry, with $2\sum_{r=1}^k \vl_r\leq\nc$,
as was explained in Section 3.2.
In the dual theory, we also have a massive adjoint field, and the
analysis proceeds similarly. However, we need to again consider stability.
It is convenient to write the symmetry group of the magnetic theory in
the form:
$$SO([2k+1]\nf-\nc+4)\rightarrow
SO\left(\nf-\nc+4+2\sum_{r=1}^k j_r\right)\times
U(\nf-j_1)\times\ldots\times U(\nf-j_k) \ .$$
 Stability requires $j_r\geq 0$ and
$\nf+4\geq\nf-\nc+4+2\sum_{r=1}^k j_r$,
and so the stable ground states are in one-to-one
correspondence with those of the electric theory.  Identifying $p_r=j_r$
makes it clear that the duality of \refs{\NAD, \somods} is maintained.

One may also consider deformations by $\tr{\ad^r}$ for $r>2$.
The analysis is similar to the $Sp(2\nc)$ case studied earlier
and we omit the details. The duality map
reduces in all cases in the appropriate fashion.

\newsec{Comments and Conclusions}

Hints of relations between N=2 duality and that of N=1 have been noted
by several authors \refs{\NAD, \emop, \somods, \kutschw}.
In \emop\ we conjectured a specific relation
between the two (which has since been verified \spa)
that explains the appearance of the singlet
fields $M$ under N=1 duality.    This relationship appears to be present in
the models of \KUT\ and \kutschw\ and persists in the models
of this paper; however the models of \keni\ appear to
show that the duality of
N=1 models is not restricted to those which can be derived from N=2
theories.

Still, a general pattern has emerged.  All of the known
classes of models contain
certain special theories whose duals have the same gauge group.
The key feature of these self-dual theories is that they
have marginal operators and associated lines of conformal
fixed points \emop\ which connect the electric
theory to a theory which is isomorphic, up to a reflection in the flavor
group, to the magnetic theory.  This phenomenon is exactly that
of certain N=2 finite models, where the marginal coupling constant is
the gauge coupling; the electric theory at $g$ is dual to the
magnetic theory at $1/g$, which itself is isomorphic (up to
a flavor group reflection) to the electric theory at $1/g$.

The models of Kutasov and Schwimmer \refs{\KUT, \kutschw}
have several properties
which relate them to N=2.\foot{Many of the observations made
below were also made in \kutschw.}
The self-dual points in the $SU(\nc)$ theories
are found by beginning with the finite N=2 theory with $2\nc$
hypermultiplets, whose gauge coupling is marginal.
Using N=2 violating mass terms to remove
all but $1/k$ of the quarks, a low energy theory is
generated with a marginal superpotential \emop\
of the form $W=hQ_rX^{k}\tQ^r$.  The relevant perturbation by
$\Tr X^{k+1}$ apparently drives the theory to a new fixed point with a
marginal superpotential
$W = {\lambda\over k+1}\Tr X^{k+1} + hQ^rX^{k}\tQ_r$.
In the small $h$ limit the model becomes the electric theory of
\kutschw.  The equation of motion for $X$, multiplied by $Q^s$
and $\tQ_s$, gives
$$
hQ^rX^{k}\tQ_r =
-{h^2\over\lambda}\sum_{n=0}^{k-1}(Q^rX^{j}\tQ_s)(Q^rX^{k-j-1}\tQ_s)
                           + \cdots
$$
This suggests that one should introduce auxiliary mesons as in
\emop\ and replace this operator with
$$
\sum_{n=0}^{k-1} \left[(N_j)_r^s(Q^rX^{j}\tQ_s) +
{\lambda\over 4 h^2}(N_j)_r^s (N_{k-j-1})^r_s\right]
                           + \cdots
$$
Thus the large $h$ limit, for which the $N$'s are massless, yields a model
which is (roughly) isomorphic to the magnetic theory of \kutschw, up to the
usual reflection of the flavor group.  For $k=1$ this argument
is consistent with \emop.  However, we do not as yet see how to
turn this sketch into a rigorous derivation.

Nonetheless, these observations served as motivation to find the theories
of the present paper, which again can be obtained by integrating
out $1/(2k+1)$ of the hypermultiplets of an N=2 model, perturbing
the theory by $\Tr X^{2(k+1)}$, and using the equation of motion
as a guide to find the magnetic theory at the self-dual point.
The mapping of operators is straightforward, except for baryons
in $SO(\nc)$ for which the N=2
duality transformation is still unknown.  The models of \keni\
are not derivable from N=2 theories, but they have the same structure:
the self-dual models have marginal operators that connect the
electric theory to a model isomorphic to the magnetic theory.

Other comments made at the end of \kutschw\ also apply to our models.
Unitarity of the theory requires certain operators to decouple; it
is unclear how this occurs.  The dimensions of operators in the
theory without a superpotential are unknown; from the theories
studied here, one may infer certain aspects of the behavior
of $\Tr X^r$ as a function of $\nf$ and $\nc$.  It would also be
interesting to study the presumably intricate behavior of the $SO(\nc)$
models, many of which have dyonic as well as magnetic duals, which
we have not addressed in this paper.  Finally, let us note that
the theory with $\nf=0$ is a theory of great interest since
one can reach it from a pure N=2 Yang-Mills theory.

It is evident that the models studied in
Refs.~\refs{\NAD, \KUT, \somods, \kutschw,\kenpoul,\keni},
along with those of this paper, all lie in the same
class.  Certain patterns are beginning to become apparent, and
we hope to study them further in the near future.

\bigskip

\centerline{{\bf Acknowledgments}}

We thank Ken Intriligator for interesting discussions.
R.G.L.~thanks the CERN Theory Division and M.J.S.~thanks the theory
group at Case Western Reserve University for their hospitality during the
completion of this paper.
This work was supported in part by DOE grant
\#DE-FG05-90ER40559.

\listrefs\end